# A Study of Knowledge Sharing related to Covid-19 Pandemic in Stack Overflow

*Konstantinos Georgiou*
School of Informatics
Aristotle University of Thessaloniki
Thessaloniki, Greece
kageorgiou@csd.auth.gr

*Nikolaos Mittas*
Department of Chemistry
International Hellenic University, Kavala, Greece
nmittas@chem.ihu.gr

*Lefteris Angelis*
School of Informatics
Aristotle University of Thessaloniki
Thessaloniki, Greece
lef@csd.auth.gr

*Alexander Chatzigeorgiou*
Department of Applied Informatics
University of Macedonia, Thessaloniki, Greece
achat@uom.edu.gr

*Abstract*— The Covid-19 outbreak, beyond its tragic effects, has changed to an unprecedented extent almost every aspect of human activity throughout the world. At the same time, the pandemic has stimulated enormous amount of research by scientists across various disciplines, seeking to study the phenomenon itself, its epidemiological characteristics and ways to confront its consequences. Information Technology, and particularly Data Science, drive innovation in all related to Covid-19 biomedical fields. Acknowledging that software developers routinely resort to open 'question & answer' communities like Stack Overflow to seek advice on solving technical issues, we have performed an empirical study to investigate the extent, evolution and characteristics of Covid-19 related posts. In particular, through the study of 464 Stack Overflow questions posted mainly in February and March 2020 and leveraging the power of text mining, we attempt to shed light into the interest of developers in Covid-19 related topics and the most popular technological problems for which the users seek information. The findings reveal that indeed this global crisis sparked off an intense and increasing activity in Stack Overflow with most post topics reflecting a strong interest on the analysis of Covid-19 data, primarily using Python technologies.

*Keywords—Covid-19; Stack Overflow; Data Analytics*

## I. INTRODUCTION

The recent outbreak of the novel coronavirus disease (Covid-19) that has been initially detected in Wuhan, China and has rapidly spread all over the world, has caused a global crisis. Besides the profound impact in health care, the crisis has expanded to all other aspects of everyday life, including social relations and economic activities. A unique characteristic of the crisis is its rapid evolvement, i.e. the situation changes from day to day. Due to the growing number of infection epicenters, governments have put emphasis on tight restrictive measures that prohibit access to work, travels, meetings, etc. Such strategies essentially at first encouraged and then necessitated distance work, education, entertainment and social activities. These measures soon led to increased and massive demand of technological support, especially regarding communications, awareness, adaptation of e-government and e-shopping systems, etc.

Apart from technologies supporting everyday activities during the Covid-19 outbreak, there are growing needs for Information Technologies (IT) that could aid the battle against the disease. Similarly, researchers all over the world are attempting to investigate the root causes of the phenomenon, its spread patterns and mathematical models governing its evolution. For example, data science methodologies (e.g. data storing and management) are particularly important for epidemiologists and for all those taking decisions based on the distribution and the mortality rates of the disease. Furthermore, other areas such as bioinformatics and biostatistics are used in genomic and clinical research, whereas, even for information purposes, models and visualization are relying on data analytics.

The motivation of our work is based on our own observations and our intuitive perception of the IT activities in demand during the outbreak. Therefore, we decided to investigate whether and how the new intense circumstances are reflected on relevant developers' posts in a "Q&A" portal such as *Stack Overflow* (SO). The goal is to find, retrieve and study SO posts that are related to Covid-19 so as to understand why and for which topics developers are interested and how these topics are associated. Also, our study aims to investigate the evolution of posts in a time frame where the crisis is continuously escalating. We should note that with the term 'developers' we do not necessarily assume that users posting on SO are software professionals. Especially with respect to Covid-19, we consider it highly possible that scientists from various domains (e.g. medical sciences, biostatistics, public health scientists, etc.) are pursuing research that relies on *scientific software*[1] [1] mainly targeting collection, analysis, visualization and interpretation of Covid-19 related data.

We believe that this research is useful to a wide range of scientists from different disciplines since it contributes in understanding the interconnections of real world disasters and crises in an era of massive available information, especially within communities that employ IT to access such data sources and consume them.

## II. RELATED WORK

SO is continuously attracting the interest of researchers, as a repository of technological knowledge and expertise. Thus, many attempts have been made at tracking its evolution and pinpointing the areas of interest it includes.

---

[1] We refer to software created by scientists and engineers as *scientific software* following the definition by Heaton and Carver

Attempts focusing at the website's community structure [2] analyze the answers to questions and, by linking co-answerers and answer votes, predict the quality of questions and their technological content. By inferring the community structure through the analysis of posts, affiliation networks have also been employed [3] to track user activity and cluster common tags into thematically similar groups. Research has also been performed in the analysis of tags and their use by developers. Topic shifting [4] and topic modelling [5] are used in combination with the time factor to identify trend changes in tag usage along with the characteristics that render a tag suitable for a question.

Regarding the nature of the questions posted in multiple categories such as mobile development [6], security [7] or questions of all types, results indicate that although most questions are mainly inquisitive (of the types "what", "how" and "why"), there are questions of more technical nature, associated with specific fields of specialization [8,9].

Graph based approaches have been also used to represent networks of tag co-occurrences [10] and tag-related communities [11,12]. Finally, aspects like the personality or the emotions of developers that post questions or answers is an active topic of research, especially with respect to their reputation in the community [13].

## III. RESEARCH OBJECTIVES AND RESEARCH QUESTIONS

The main idea behind this study is to investigate, whether Covid-19 outbreak has affected the activity of developers' triggering them to post questions related to coronavirus topics in knowledge sharing communities. To this regard, we focus solely on posts, where the problem/issue being discussed is associated to the adaption of IT methodologies (e.g. data science, software engineering etc.) aiming to investigate, understand and provide solutions to Covid-19 health crisis. Hence, we do not take into consideration posts, where the topic of discussion concerns for example the use of collaboration technologies that are extensively used by organizations, educational institutions etc. in order to facilitate their operational activities during the coronavirus lockdown.

Figure 1 Example of a coronavirus-related post

In order to better illustrate our main objective, we provide a representative example of what we are referring to as a coronavirus-related post (Figure 1). The title of the post summarizes that the main topic of the question being asked by the user is to extract a dataset of Covid-19 cases comprising information regarding confirmed cases, country of interest, deaths etc. as it is concluded by the inspection of the code snippet of the post. Even though we cannot be absolutely certain regarding the poster's intentions, the question is certainly associated to technological challenges emerging from the Covid-19 pandemic. Except for the title and the body of the question, the four tags of this specific post ("python", "api", "web-scrapping" and "beautifulsoup") directly point to the context of the discussed topic, which, in this case, is the broad scientific domain of data science.

Based on the previous clarifications, we formulate the following research questions:

**[RQ$_{1a}$]** *Are developers concerned about coronavirus-related topics in SO?*

**[RQ$_{1b}$]** *When did this interest arise and which is the evolution trend over the examined period?*

*Motivation*: Given the severity of the crisis and its detrimental effects to all aspects of our daily lives, RQ$_{1a}$ aims to explore, whether developers have been triggered to investigate and get involved into specific challenges related to Covid-19 Pandemic, whereas in RQ$_{1b}$, we focus on the examination of the evolement of this phenomenon from its start and thorough the examined time period of this study.

**[RQ$_2$]** *Which are the characteristics of the coronavirus-related posts?*

*Motivation*: Since Covid-19 outbreak is an emerging, rapidly evolving situation, there is no prior knowledge about the sharing of knowledge and the related posts that can be found in SO. To this regard, there is a need to explore and acquire knowledge about the characteristics of the coronavirus-related posts.

**[RQ$_{3a}$]** *Which are the most popular problem categories expressed by SO tags among Covid-19-related posts?*

**[RQ$_{3b}$]** *How these problem categories are associated to each other?*

*Motivation*: Given the rapid interest on coronavirus-related research challenges, in RQ$_{3a}$, we focus on the identification of the most popular technological aspects that are discussed by developers. Each post is labelled with one to five tags defined by the user in order to provide a general overview of the question by associating it to certain technological content as the basis for future discussion [8]. To this regard, Beyer et al. [8] point out that SO posts can be classified into two question categories: (*i*) *problems* and (*ii*) *questions*. Problems, expressed by SO tags, refer to the topics or technologies that are discussed, whereas questions are associated to the purpose of the post representing "*the kind of information requested in a way that is orthogonal to any particular technology*" [9].

Hence, in RQ$_{3a}$, we wish to investigate the technological content of Covid-19 related topics, and for this reason, we extract and analyse the SO tags associated to each post. In RQ$_{3b}$, our aim is to explore underlying patterns expressed by the co-occurrences of tags assuming that they contain information about the interconnected technological aspects discussed in the frame of coronavirus-related topics.

**[RQ4]** *What types of topics are developers asking about?*
*Motivation*: In RQ4, we focus on the discovery of what is being asked in coronavirus-related posts. Although tags can provide meaningful insights of technological content, they do not identify purposes, issues, problems and generally the reasons that lead the users to post a question [8]. To this end, we make use of a topic modelling methodology on the corpus of the title text in order to discover what are the main topics of discussion for coronavirus-related posts.

## IV. METHODOLOGY

In this Section, we provide details regarding the methodology of our experimental setup, which is illustrated in Figure 2 and can be described as seven-step process consisting of (*i*) data collection, (*ii*) feature extraction, (*iii*) data cleaning and pre-processing, (*iv*) data representation, (*v*) data analytics, (*vi*) knowledge synthesis and (*vii*) dissemination of the findings.

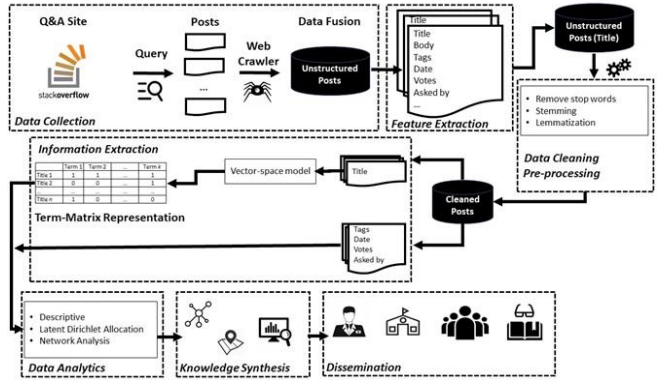

Figure 2 Methodology of the study

### A. Data Collection

Driven by the main objective and the RQs (Section 3), we decided to collect posts found in SO, a well-known Q&A forum. SO is popular for knowledge sharing among software development due to its timeliness, in the sense that it reflects quickly any emerging technological trends [14]. Since our main objective is the investigation of what developers are asking about coronavirus-related topics, we must note that the data collection is restricted only to question posts. The search string that was used was quite broad, comprising synonyms of coronavirus (*"coronavirus"* OR *"covid*"* OR *"corona-virus"*) to retrieve the entire set of relevant questions posted until the 1st of April, which was the final date of collection. The application of the search string on the SO search engine returned 601 candidate posts in the form of semi-structured web documents. Document collection was achieved by using a web scraper built in Python based on the Selenium package[2]. The scraper retrieved all questions on each page and iterated each question individually to collect additional features not found on the question's display on the results page.

---
[2] selenium. Retrieved from https://pypi.org/project/selenium/

### B. Feature Extraction

The basic unit of analysis is the *question post* (or simply post), which is a web document composed of a number of fields with a variety of entries. Generally, a post $P$ can be defined as a tuple of a finite number of ordered elements of the general form

$$P = \langle id, t, b, tag, d, na, u \rangle \quad (1)$$

where $id$ is the *identification number*, $t$ is the *title*, $b$ is the *body*, $tag$ are the associated *tags*, $d$ is the *posting date*, $na$ is the *number of received answers*, $u$ is the *identification number of the post owner*.

### C. Data Cleaning and Pre-processing

As we have already mentioned, the tuples of posts comprise of both unstructured (i.e. *title*, *body*) and structured (i.e. *date*, *number of received answers* etc.) features. Regarding unstructured text, *Text Mining* (TM) techniques were applied in order to reduce the size and noise of the collected data through *data cleaning* and *text pre-processing* steps. Specifically, we first filter posts that are irrelevant to the general objectives of the current study, i.e. posts including messages of support, information related to technological assistance due to coronavirus lockdown etc. After this cleaning process, we conduct the following steps on the remaining 464 posts: (*i*) removal of stop-words, (*ii*) tokenization, (*iii*) punctuation removal and (*iv*) lemmatization. Regarding the list of tags, each tag was transformed into a dichotomous variable (FALSE/TRUE) based on its absence/presence in a post.

### D. Data Analytics

The next step involves the application of data analytics methodologies on the set of features to derive conclusions about the posed RQs (Section 3). Regarding RQ1a and RQ1b, we study the distribution of the collected posts during the examined period our study, whereas for RQ2, we conduct a kind of descriptive statistics analysis on the extracted features of posts.

Regarding the analysis on the list of tags, we initially explore the number of tags for each post (RQ3a) and then, we make use of graph theory to examine whether there are inter-connections among a subset of tags (RQ3b). More specifically, we adopt a similar approach proposed by Cui et al. [13] that is used for the examination of the dynamic trends in collaborative tagging systems. According to this approach, a *tagging action* can be described as a tuple of three ordered elements:

$$tagging = \langle u_i, p_j, tag_k \rangle \quad (2)$$

implying that the user $i$ tagged post $j$ with a set of tags $k$. Based on this definition, we develop an *Association Rule Graph* (ARG) [13] for tags and their associations to represent inter-connected technological aspects.

Three concepts play pivotal role to the ARG approach: The *frequency* ($freq(tag_i)$) of a tag, defined as the number of occurrences of a tag $i$ in the set of the collected posts, the *support* ($supp(tag_i, tag_j)$), defined as the number of co-occurrences of a specific pair of tags $\{tag_i, tag_j\}$ in the set of the collected posts and the *confidence* ($conf(tag_i \rightarrow tag_j)$) representing the conditional probability of the occurrence of $tag_j$ in a post, given that the post is already tagged by $tag_i$, where $freq(tag_i) < freq(tag_j)$. The *confidence* is defined as the ratio of the support for a given pair of $\{tag_i, tag_j\}$ to the frequency of the tag with the lower frequency ($tag_i$).

$$conf(tag_i \rightarrow tag_j) = \frac{supp(tag_i, tag_j)}{freq(tag_i)} \quad (3)$$

Based on the previous definitions, ARG is defined as a directed graph $G = (V, E)$, where $V$ and $E$ are the set of vertices and edges, respectively. In our case, each tag ($tag_i$) from the set of distinct tags represents a vertex accompanied by a weight ($w_{tag_i}$) evaluated by the tag frequency ($freq(tag_i)$) of this specific tag. In addition, for each pair of tags $\{tag_i, tag_j\}$ that co-occurred in the set of the collection of posts and satisfying and $freq(tag_i) < freq(tag_j)$, there is a directed edge accompanied by an edge weight computed by the confidence ($conf(tag_i \rightarrow tag_j)$) for this specific pair.

For RQ4, we perform *Latent Dirichlet Allocation* (LDA) analysis on the question titles, since titles provide straightforward information about the problem being asked by the user. LDA [15] is a well-known probabilistic model for topic extraction in a collection of documents that has been extensively used in many experimental studies for topic modeling on textual content of SO posts [9]. The main idea behind LDA is that documents are represented as probability distributions over latent topics, where each topic is represented by a distribution of words [15]. Thus, LDA unveils the hidden topics by taking into consideration the patterns of the words that tend to co-occur frequently in the set of documents of the corpus. The selection of the number of topics $K$ is based on experimentation with different values of $K$, since it is a user-defined parameter [8] and there is not an optimum value for all experimental setups [14]. Generally, a high value of $K$ results to the identification of a detailed collection of topics, whereas smaller values of $K$ result to more general collection of topics [6, 16].

Based on the main scope of RQ4, the total number of posts and the restricted number of bag-of-words that can be found in the title of each post, we decided to set $K = 6$, since this parameter value provided a broad collection of topics capturing the general patterns that are hidden in the collection of posts through the implementation of the LDA model provided by the Gensim Python package[3] on the corpus consisting of all post titles.

V. RESULTS

In this section, we present the findings of this study based on the posed RQs.

---
[3] gensim: topic modelling for humans.
https://radimrehurek.com/gensim/

A. *[RQ1a] Are developers concerned about coronavirus-related topics in SO?*
*[RQ1b] When did this interest arise, and which is the evolution trend over the examined period?*

The search in SO indicates that "post-zero" is traced back to January 26, in which the user wished to print multiple identical values from an array of Covid-19-related text in PHP. After the cleaning step described in Section 4, the final dataset contains 464 retrieved Covid-19 related posts that constitute the basis of our study.

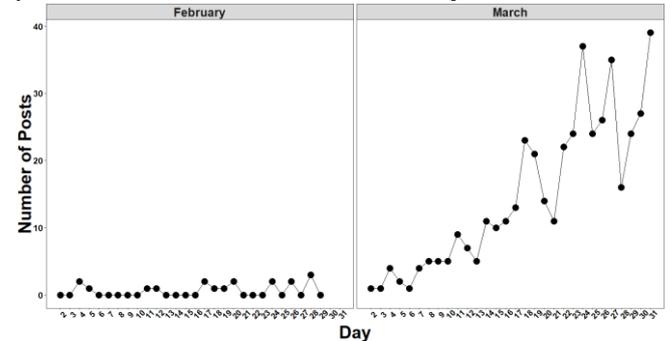

Figure 3 Number of Covid-19 related posts for each day of February/March

In Figure 3, we present the number of posts per day for two specific months (February and March) but we have also to note that the final dataset contains also few cases posted on January (including "post-zero") and April 1st (i.e. the final day of the data collection process). The figure shows a significant increase of posts for month March (approximately 95% of all posts) compared to month February (approximately 4% of all posts), when the impacts and consequences of the Covid-19 outbreak presented an explosive spread to healthcare systems, businesses, financial markets and economies and generally to our society.

B. *[RQ2] Which are the characteristics of the coronavirus-related posts?*

In order to gain better insights about the characteristics of the observed phenomenon, we initially performed statistical analysis on the associated features that can be found in each post.

Figure 4a presents the distribution of the number of received answers for the set of posts indicating that more than 60% of the posts has already been responded by at least one user (mean value $M = 0.87$, standard deviation $SD = 0.88$, median $Mdn = 1$). Considering that we have tracked Covid-19 related posts for a two-month period, it is rather reasonable that most posts have received a limited number of answers. Due to the observed growing interest on coronavirus-related, there is a need to examine, whether these questions are posted by different users or whether, there is only a specific proportion of users that are interested in. To this regard, we evaluate the percentage of unique users who posted a question related to a coronavirus issue. The percentage of unique users (# of different users posted one question divided by the # of posts) was 79.09% signifying a broad awareness of the community.

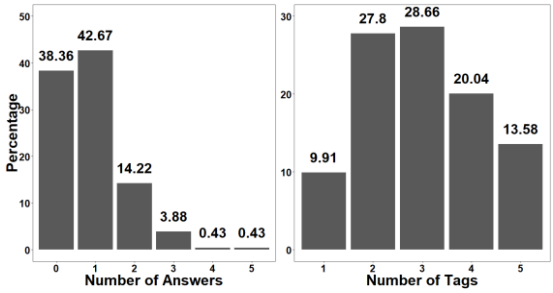

Figure 4 Percentage distribution of the number of (a) answers and (b) tags

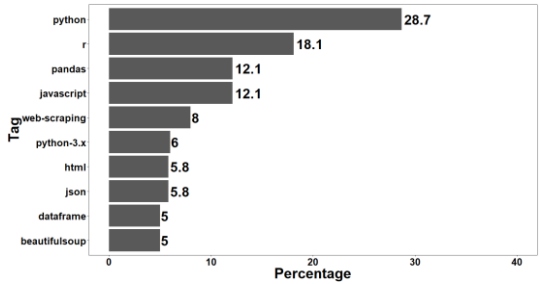

Figure 5 Percentage distribution of the top 10 popular tags

C. *[RQ3<sub>a</sub>] Which are the most popular problem categories expressed by SO tags among coronavirus-related topics?*

*[RQ3<sub>b</sub>] How are these problem categories associated to each other?*

The analysis on the occurrences of SO tags showed that most users (56.46%) defined from two up to three tags (Figure 4b). Figure 5 presents the distribution of the 10 most popular tags from the set of 482 total tags, from which we can observe that the tags "python" and "r", which are two of the most known programming languages for Data Science, significantly dominate in terms of their occurrences.

We have also to note that the percentages refer to the identified distinct tags found in the posts, which practically means that these percentages may be significantly higher, if we take into consideration the fact that for any tag, there may be a related tag or a synonym with the same meaning. For example, the sixth most popular tag presented in the dataset is "python-3.x", which is in fact, a tag that is used for posts "*about Python programming language that are specific to version 3+ of the language*"[4]. To this end, the total percentage of posts that is associated to Python programming language after summing up other unique tags referring to different versions of this specific language (i.e. "python-2.7", "python-3.7", "python-3.x") is 35.56%.

Regarding the other tags presented in the list of the top popular ones, an interesting conclusion is the fact that they are also related to Data Science. We have also to keep in mind that Data Science is all about using data in order to gain new insights that you would otherwise have missed [17]. It is an inter-disciplinary field comprising scientific methods and processes for the collection, organization, processing, storage, analysis and visualization of data with the aim of extracting knowledge hidden in unstructured, semi-structured and structured data [18]. Due to the multifaceted nature of Data Science and the need for interdisciplinary competences and skills, there is a variety of challenges and tasks that a practitioner should be able to handle. The findings suggest that the remaining most popular tags (except from "python" and "r") are closely related to phases of the whole lifecycle of Data Science. For example, the tag "web-scrapping" and "json" are related to the collection phase through the extraction of information from websites, whereas "beautifulsoup" is a Python package that is used for parsing web documents. In addition, "dataframe" is a well-known data structure in several languages (e.g. Python, R etc.) for the organization of information in data analytics and this is also the case for "pandas", which is a Python library for data manipulation.

Finally, the posts tagged by "html" or "javascript" reveal an interest to seek knowledge about scraping content from HTML documents (usually residing in Javascript-enabled pages), most often through parsers developed in Python or node.js or with the help of test automation tools. It becomes evident that once again the focus is on data collection from online sources and the developers of scientific software wish to automate the parsing from websites with dynamic content and convert the extracted information into data structures of file formats for further processing.

Regarding the topic of discussion, the manual review process through the inspection of the title and body of the posts reveals that the users are mostly interested in collecting, organizing and analyzing data of Covid-19 global cases (i.e. confirmed, deaths, recovered etc.).

After the identification of the most popular tags found in coronavirus-related topics (RQ3a), the focus is now on the investigation of the interdependencies between technological aspects (RQ3b). To this regard, we construct an ARG for the top 10 popular tags as described in Section IV. To derive meaningful conclusions avoiding the problem of presenting noise nodes and edges, we decided to exclude association rules with low frequencies of the antecedent and generally high values for confidence.

The inspection of Figure 6 indicates that the two tags of the programming languages "python" and "r" are represented by generally large circles (the area of the circle represents the frequency of the tag, the direction of an edge $(tag_i \rightarrow tag_j)$ is defined by the constraint $freq(tag_i) < freq(tag_j)$, whereas the width of the edge is weighted according to the value of confidence $(conf(tag_i \rightarrow tag_j))$. Several well-known technologies are associated to the tag "python", including libraries and frameworks for web-parsing ("beautifulsoup", "selenium"), data manipulation ("pandas"), interactive visualization ("matplotlib", "plotly", "seaborn", "choropleth") and geospatial data analysis ("geopandas", "geojson"). Finally, the tag "python" is also related to more general tags such as "data-science", "dataframe", "dataset", "curve-fitting" etc.

---

[4] https://stackoverflow.com/questions/tagged/python-3.x

With respect to another frequent tag, namely "javascript", we observe that almost all associated nodes point to libraries, frameworks and language elements needed for extracting content or posting from/to the Web: the widely used libraries "jQuery" and "d3.js", are related to HTML/XML document manipulation, while module "cheerio", method "get" and API "fetch" facilitate the manipulation of "http" requests and responses. From these tags, it is confirmed that developers of scientific software in the era of Covid-19 are interested in getting content from sites posting relevant information (presumably virus-related demographics). The nodes corresponding to the API "google-maps" and the library "leaflet" express the interest to create and annotate maps with custom content using Javascript (as for example in case of marking virus hotspots). At the same time the API "google-sheet", method "filter" and questions related to "for-loop" instructions are associated with attempts to process the data extracted from the web and store it into arrays of spreadsheets. Finally, following recent trends, developers are posting questions on state-of-practice front-end frameworks such as "react.js" and "vue.js" but are also interested in using "node.js" for server-side scripting, in the context of setting up the user interface and the back-end of Covid-19 related applications.

*D. [RQ$_4$] What types of topics are developers asking about?*

The results of the application of LDA on the corpus of titles are summarized in Table 1 presenting the most frequent words for each one of the six extracted topics. Moreover, we tried to reify the conceptual meaning of each latent topic by providing a description of topics developers are asking about coronavirus-related posts.

An interesting finding derived from the inspection of the list of frequent words is the fact that the most common type of question is "*how to ...*", a type generally found frequently in SO posts [9,18,19,6,16,8]. Furthermore, the word "*error*" designates that developers frequently face problems related to occurrences of errors in their code.

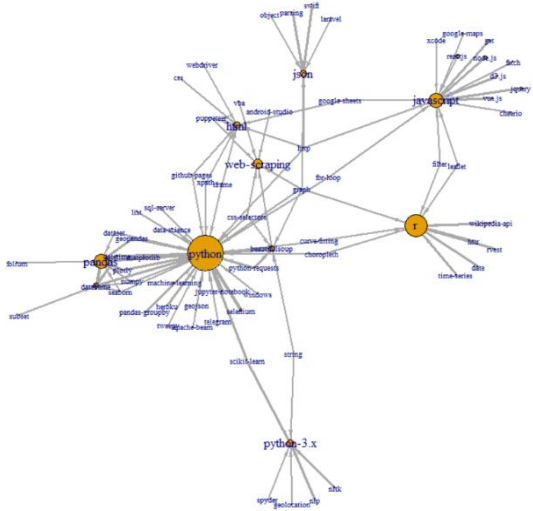

Figure 6 Association rule graph for the top 10 popular tags

Besides the type of the question, the frequent words provide to us a more detailed overview of the content of the problem being discussed compared to tags. Again, it is clear that the interest is focused on technological challenges related to phases of Data Science lifecycle and to more specific associated technologies such as specific programming languages. The most popular technological topic seems to be the collection of data related to Covid-19 cases through web-scraping techniques and content retrieval from websites, whereas the users also post questions related to the data analysis and visualization of reported cases through maps for geospatial analysis. Finally, as far as specific technologies concern, it is evident again that languages Python and R play a predominant role to the community of developers for examining challenges related to Covid-19 pandemic.

| Topic | Frequent Words |
|---|---|
| *Web-scrapping and data manipulation with Python* | how, python, pandas, value, column, list, string, web, table, scraping |
| *Error messages in importing data in R* | how, data, error, when, r, object, csv, value, file |
| *Importing web content to data structures and files* | how, using, data, table, file, get, website, name |
| *Importing web content to data structures and files in R* | how, data, error, using, file, column, r, json, url, api |
| *Data manipulation and analysis (time series) in R* | how, data, pandas, error, when, using, time, r, value, api |
| *Data visualization (time-series, geospatial) in R* | how, using, data, r, date, api, website, map, problem, value |

Table 1. Topic modeling of Covid-19 related posts

VI. DISCUSSION

In this section, we discuss the key findings of the study organized per research question and the implications to stakeholders.

RQ$_1$: *Developers interest on Covid-19-related topics in SO*

From the number of Covid-19 related posts for each day of February and March 2020 it becomes clearly evident that the developers' increasing interest over time follows closely the global trends for the number of cases, fatalities and even the popularity of Covid-19 terms in search engines. This is reasonable but also proves the quick responsiveness of the scientific software community against global phenomena and emerging topics and the pressing needs for multiple representations of data. It is also interesting that a periodicity in the downward dents in the number of posts can be observed: these drops seem to coincide with the end of the corresponding weeks, as they mostly occur on Fridays and Saturdays. Further investigation can reveal whether the observed trend in the number of posts will match the Covid-19 pandemic for the upcoming months.

RQ$_2$: *Characteristics of the coronavirus-related posts*

Despite the rapidly increasing trend in the number of Covid-19 related posts, it seems that the pandemic has taken the world by storm. All identified Covid-19 posts have received a rather limited number of answers (usually one answer) and at the same time most users posting Covid-19 related questions are unique. This is of course a consequence of the limited period in which the analysis of SO posts has been carried out, but it could also imply that the rapid developments did not allow the forming of mature research communities. In other words, as in any other emerging crises or trends, the first researchers who attempt to analyze and interpret the phenomenon are working independently and loosely interact with each other.

RQ$_3$: *Popular problem categories expressed by SO tags*

The analysis of the identified SO tags clearly implies a strong interest on seeking knowledge about the collection, organization, processing, storage, analysis and visualization of data. The developers' goal is in most of the cases to extract information from web pages and the primary languages for this purpose appear to be Python and R. Considering that all posts are related to Covid-19, we can postulate that developers of scientific software from various domains focus on Data Science techniques to analyze the tremendous dimensions of the coronavirus outbreak. Thus, knowledge sharing communities such as SO prove useful not only for solving everyday technical challenges faced by software professionals but also as a means of collaboration and essential knowledge dissemination in cases of emergencies or global crises. It should also be praised that developers are willing to share the knowledge they hold to assist fellow researchers underlining the benefits of open source projects and communities.

RQ$_4$: *Topics of interest*

The application of LDA on the corpus of post titles confirm the overall picture obtained by RQ$_3$. The extracted topics reveal that during the Covid-19 pandemic developers of scientific software are applying data science techniques to a typical data science problem, that of analyzing and interpreting the day-by-day changing virus-related figures. Developers experiment with collecting data mostly through web-scraping and in most of the cases build python, r, or javascript applications to visualize aspects of the phenomenon. The most frequent words associated with each topic essentially reflect the early stages of programming in a new environment where "how-to" questions come first and fixing errors right after. The pattern of words possibly implies that beyond the experienced developers, researchers with less experience in programming have also been attracted by the gigantic dimensions of the crisis and have attempted to write their own applications.

All findings provide evidence that the outbreak of a global health crisis triggers increased interest for data collection and processing from software developers. Although it is difficult to infer whether the interest is professional or personal, the trend is clear. Based on the evidences, we can assume as main motives the need for understanding the nature of the crisis and the need for forecasting its evolvement. We believe that these findings are interesting and useful for all stakeholders involved in software development (web sites, social awareness applications, governmental services etc) either individuals or organizations, since they provide an overall picture of the technologies that are related to the needs and trends of a crisis. Moreover, stakeholders involved in areas directly or indirectly related to technology, such as financial forecasts, biomedical research and business management are potentially interested in the findings of the current study.

## VII. THREATS TO VALIDITY

In this section, we present and discuss potential threats to the validity of study. Regarding the internal validity, the identification of posts related to Covid-19 topics was performed by an automated process through the search engine of SO. The search strings were quite broad, since we only included general terms related to Covid-19, so as to identify a high proportion of relevant posts but there is always the risk of omitting questions, in which a different terminology for the pandemic might have been used. In addition, we have removed non-relevant posts, containing the keywords of the search string but addressing a problem that is associated to a general topic resulting from the coronavirus lockdown (e.g. issues related to the utilization of collaborative technologies). To mitigate bias from this action, filtering was performed by the first and the second author independently and by discussing conflicts. Moreover, the collection, pre-processing and analysis of data were based on mature packages of Python and R.

Regarding the decision on the number of topics which is an important input parameter for the extraction of topics through LDA, the selection of the optimal number is not a trivial process. For this reason, we conducted several experiments utilizing different values for the number of topics having in mind the general scope of the posed research question that is to investigate the broad topics of discussion for coronavirus-related posts. With respect to the interpretation of the extracted topics, the process is solely based on the analysis conducted by reading the associated terms, since LDA results to the identification of latent concepts that should be reified by the practitioners of each scientific domain.

Regarding external validity, one potential threat is the fact that we restricted the analysis and inferential process on questions posted in SO. Although SO is one of the most popular Q&A sites for software developers, there are opportunities for further research in order to compare and generalize our findings to other knowledge sharing forums. Additionally, we have to note the limited timeframe of the collection phase resulting to the retrieval of a specific number of posts, which is certainly smaller compared to other studies investigating other more general topics. However, we have to take into consideration that Covid-19 pandemic is an unexpected and emerging situation with detrimental effects to any aspect of human activity and for this reason, it triggers the adoption of a variety of necessitated strategies and actions to understand and mitigate the consequences of this phenomenon, even in their infantile form. Finally, the extraction of topics of discussion was based on the analysis conducted on the titles of posts. Despite the fact that the title of a question is expected to provide a general overview of the problem being asked, the inclusion of the corpus found in the body may improve the inferential process, even if it may also introduce some noise.

## VIII. CONCLUSIONS AND FUTURE WORK

Covid-19 is considered by many the worst health crisis of a generation that is challenging the world. For every society the top priority has been the minimization of spread and the treatment of infected people. Within the medical approaches in the 'arsenal of health' to cope with the pandemic, the paramount importance of epidemiology has become clear. Proper policy decisions can be taken only if we understand and predict the distribution, growth patterns and spread mechanism of the disease. To this end, Data Science and Information Technologies have a lot to offer.

In this study, we have focused on the interest of scientific software developers on Covid-19 related topics based on SO posts to shed light into their concerns, goals and means. The results indicate an increasing interest that matches the evolution of the pandemic; however, the development of software to analyze the phenomenon has just started. With respect to the information that developers seek, the findings clearly indicate the ongoing effort to collect, store, analyze, visualize and interpret Covid-19 related data. Collection is mostly performed through Web scraping, with Python, R and javascript clearly being the languages of choice.

The amount of scientific activity on challenges pertaining to Covid-19 is expected to boom in the next few months. Considering that all kinds of scientific studies rely on the analysis of massive Covid-19 related information, we also anticipate a tremendous increase in associated software development activity. Thus, we plan to extend this study by considering larger datasets of SO posts but also more sources, such as open-source software repositories which are expected to host relevant applications.